\documentclass[]{rmaa}
\usepackage{natbib}
\bibpunct{(}{)}{;}{a}{}{,}

\usepackage{paralist}

\usepackage{psfrag,color}

\title{Shadows and Photoevaporated Flows from Neutral \\ Clumps Exposed
to Two Ionizing Sources} 

\author{
  A. H. Cerqueira,\altaffilmark{1,2} 
  J. Cant\'o,\altaffilmark{3}
  A. C. Raga \altaffilmark{2}
  and M. J. Vasconcelos \altaffilmark{1,2}}

\altaffiltext{1}{LATO-DCET, UESC, Ilh\'eus, Bahia, Brasil}
\altaffiltext{2}{Instituto de Ciencias Nucleares, UNAM, D.F., M\'exico.}
\altaffiltext{3}{Instituto de Astronom\'\i{}a, UNAM, D.F., M\'exico.}

\shortauthor{Cerqueira et al.}
\shorttitle{Shadows behind illuminated clumps}

\fulladdresses{
\item Jorge Cant\'o, Instituto de Astronom\'\i{}a, UNAM,  
  Apartado Postal 70-264, 04510, M\'exico, D.F., M\'exico.
\item Adriano H. Cerqueira and Maria Jaqueline Vasconcelos: LATO-DCET, UESC,
  Rodovia Ilh\'eus-Itabuna, km 16, Ilh\'eus, Bahia, Brasil, CEP 45662-000
  (\email{hoth, mjvasc@uesc.br}).
\item Alejandro C. Raga, Instituto de Ciencias Nucleares, UNAM,
  Apartado Postal 70-543, 04510, M\'exico, D.F., M\'exico
  (\email{raga@nucleares.unam.mx}).

}

\listofauthors{Cerqueira et al.}
\indexauthor{Cerqueira, A.H.}
\indexauthor{Cant\'o, J.}
\indexauthor{Raga, A.C.}
\indexauthor{Vasconcelos, M.J.}

\abstract{Neutral clumps immersed in HII regions are frequently found
in star formation regions. We investigate here the formation of tail of
neutral gas, which are not reached by the direct ionizing flux coming from
two massive stars, using both an analytical approximation, that allows
us to estimate the shadow geometry behind the clumps for different
initial geometric configurations, and three-dimensional numerical
simulations. We found a good agreement between both approaches to this
theoretically interesting problem. A particularly important application
could be the proplyds that are found in the Trapezium cluster in Orion,
which are being photoevaporated primordially by the O stars $\theta^1$
Ori C and $\theta^2$ Ori A.}

\resumen{Gl\'obulos neutros sumergidos en regiones HII son frecuentemente
encontrados en regiones de formaci\'on estelar. En este trabajo
investigamos la formaci\'on de regiones de sombra detr\'as de gl\'obulos
neutros iluminados por el flujo ionizante directo de dos estrellas.
Utilizamos una aproximaci\'on anal\'\i tica, la cual nos permite hacer
estimaciones de la geometr\'\i a de las sombras detr\'as de los gl\'obulos
neutros para distintas configuraciones geom\'etricas iniciales, y tambien
simulaciones num\'ericas tridimensionales. Encontramos un buen acuerdo
entre los c\'alculos anal\'\i ticos y los num\'ericos. Una aplicaci\'on
de particular relevancia pueden ser los proplyds ubicados en el c\'umulo
del Trapecio en Ori\'on, los cuales est\'an siendo fotoevaporados por
la radiaci\'on de las estrellas O  $\theta^1$ Ori C y $\theta^2$ Ori A.}

\addkeyword{hydrodynamics}
\addkeyword{H~II regions}
\addkeyword{Stars: Pre-main sequence}
\addkeyword{Stars: formation}

\begin{document}
% Typeset article header
\maketitle

\section{Introduction}
\label{intro}

High density neutral clumps immersed in a bath of photoionizing
and photodissociating radiation can be found in several regions in
our own Galaxy. To quote only a few examples, we can find cometary
globules like CG 1 in H II regions (in the Gum Nebula), for which
the major source of ultra-violet (UV) photons is the Zeta Puppis star
\citep[e.g.,][]{rei83}; the proplyds found near the Trapezium
cluster in Orion \citep[e.g.,][]{odell,bally}, as well as in other H
II regions with star formation \citep[e.g.,][]{smith}. The morphology,
dynamics and emission line spectrun of the Orion proplyds is explained as
the photoevaporation of circumstellar disks by the ionizing photon flux
that comes from $\theta^1$ Ori C \citep[][]{johns,rich}.  Another example
of high density neutral condensations in an ionizing radiation field are
Thackeray's globules in the IC 2944 region, which show a fragmented,
clumpy structure \citep{rei2}. We can also find high density clumps
associated with planetary nebulae (PN). The Helix Nebula \citep[NGC
7293, e.g.,][]{odell2} displays thousands of cometary shaped neutral
clumps, which are being photoevaporated \citep[e.g.,][]{lopez} by the
radiation field that comes from the central star. Attempts have also
been made to reproduce observational aspects of the so-called Fast Low
Ionization Emission Regions (or FLIERs) observed in a fraction of PN as
a by-product of the photoevaporation of a neutral clump by the stellar
radiation field \citep[e.g.,][]{mellema98}.

Previous analytical studies of high density neutral clumps exposed
to an ionizing photon radiation field show that these systems have two
distinct dynamical phases \citep[e.g.;][]{bertoldi1, bertoldi2,mellema98}:
an initial {\it collapse} (or {\it radiation-driven implosion}) phase
and a {\it cometary} phase. The collapse phase starts when the ionizing
photon flux reaches the clump surface, forming a D-type ionization front
(IF). This IF is preceeded by a shock front, which propagates through
the clump (at a velocity determined by the local isothermal sound speed)
compressing, heating and accelerating the clump material. The ionized gas
expands outwards (with a velocity approximately equal to the isothermal
sound speed of the ionized material) generating a photoevaporated
flow\footnote{Actually a neutral clump could be instantly ionized
without evolving through the collapse phase. See a detailed discussion
in \citet{bertoldi1}.}. The recombination of the gas downstream of the IF
(i.e., in the photoevaporated flow) partially shields the clump surface
from the ionizing photon flux that comes from the ionizing source,
and this is also an important  effect influencing the evolution
of the collapse phase. When the shock front propagates through the
entire clump, in a timescale given by $t_0 = 2R_c/v_s$ (where $R_c$
is the initial clump radius and $v_s$ is the shock speed),
the clump starts to accelerate as a whole (as a dynamical response to
the production of a photoevaporated flow; i.e., the rocket effect) and
enters the cometary phase.  The clump material is found to be completely
ionized in a timescale that depends on both $t_0$ and the shielding of
the impinging ionizing flux \citep[e.g., see][]{mellema98}.

One of the first numerical studies of an ionizing radiation field
interacting with a neutral clump was related to the problem of star
formation. \citet{klein} and \citet{sand} developed a two-dimensional
code in order to numerically reproduce the interaction of an ionizing
radiation field from (an already formed) OB star association with local
inhomogeneities in a molecular cloud. These early studies were able to
capture the radiation-driven implosion of the clump material. They have
also found that the shock induced by the ionization front at the clump
surface substantially increases the density when compared with an analytical,
one-dimensional evaluation.

The subsequent evolution of these neutral clumps to the cometary regime
was investigated numerically in two-dimensions by \citet{lef94} and
\citet{mellema98}. In particular, the numerical simulations of a cometary
globule carried out by \citet{lef94} show that, after a short timescale
($\approx 10\%$ of the clump lifetime) associated with the collapse
phase, the clump evolves to a situation of quasi-hydrostatic equilibrium,
characteristic of the cometary phase \citep[see also][]{bertoldi1}. In
a subsequent paper, \citet{lef95} show that the cometary globule CG7S
could be successfully explained as a neutral clump undergoing the
collapse phase under the influence of a nearby group of O stars. On the
other hand, \citet{mellema98} were able to reproduce the kinematic and
emission properties of FLIERs seen in PN with a model of a clump, located
in the outer parts of a PN, being photoionized by the central PN star.
They have also been able to follow the onset of the collapse phase and its
photoevaporated flow, the effect of the photoevaporated flow on the clump
shaping, and the further acceleration of the clump in the cometary phase.
More recently, three-dimensional numerical simulations of non-uniform
high density clumps \citep[][]{gon05a} subject also to the influence
of a wind \citep[][]{gon05b} have been carried out. These effects are
important for the study of proplyds as well as for Thackeray's globules.

The observed high density neutral structures associated with H II
regions and PN, project a shadow away from the direction of the impinging
ionizing photon flux. \citet{canto98} studied the shape and structure of
the shadow projected by a spherical clump in a photoionized region, also
taking into account the effect of the diffuse ionizing field. They were
able to describe (analytically and numerically) the transition between a
shadow region that is optically thin to the diffuse ionizing radiation
field to a neutral inner core inside the shadow. They also predicted
that the H$\alpha$ emission coefficient is substantially greater in the
shadow region when compared with the surrounding H II region (and, then,
to conclude that the shadow should be brighter than the H II region). A
similar approach was followed by \citet{pavlakis}, who carried out 2-D
simulations. They found that if the diffuse ionizing field is 10\% of the
direct field, the evolution of the clump is considerably different to the
evolution of the clump without the diffuse radiation field. Nevertheless,
none of these papers have attempted to follow the whole evolution of the
shadow as the clump evolves from the collapse through the cometary phases.

Observationally, some photoionized neutral structures \citep[for example,
the Orion proplyds, see][]{odell} have relatively short elongations of
neutral material into the shadow region, while others have extremely
long ``tails'' extending away from the photoionizing source. Examples of
long tails are the cometary knots in the Helix nebula \citep[see, e.g.,
the recent paper of][]{odell05} and the dark trails which cut through
the outer regions of M~42 \citep{odell00}. In these tail regions, one
has a combination of the shadowing effect and the photoionization due to
the diffuse radiation (see Cant\'o et al. 1998 and Pavlakis et al. 2001)
as well as the possible presence of a photodissociated wind coming from
the back side of the neutral clump or disk structure \citep[as has been
explored for the case of the Orion proplyds by][]{rich}.

An interesting effect is that neutral clumps in H~II regions in
some cases are being photoionized by the radiation from more than one
star. For example, the structures of some of the proplyds appear to be
affected not only by the radiation from $\theta^1$~Ori~C, but also by
$\theta2$~Ori~A (in particular the 197-427, 182-413 and 244-440 proplyds,
see O'Dell 1998 and Henney \& O'Dell 1999). A possibly more complex
example are the proplyds in the Carina nebula \citep{smith}, which are
possibly affected by the radiation from several stars in the neighborhood.
This situation was first addressed by Klein, Sanford and Whitaker (1983),
who performed two-dimensional radiation-hydrodynamics calculations of
the interaction of the radiation field from two massive stars with a
neutral clump, in order to investigate whether or not such a system
could trigger star formation within OB subgroups. The calculation
performed by Klein et al. considered two identical stars (separated by 1 pc)
and, in the middle of the straight line connecting both stars, a neutral
clump (with a radius of 0.6 pc). They found that the clump is strongly compressed
(by a factor of 170) after its implosion and reaches the
Jeans mass at some points (located in a torus that surrounds the clump
initial position) in a time scale smaller than that in which the clump
can be photo-evaporated. Thus, they conclude that stars could be formed
in OB associations by this "multiple implosion mechanism".

In the present paper, we investigate the evolution of a high density
neutral clump subject to the influence of two ionizing photon sources. In
some sense, this paper represents a generalization of the Klein et
al. (1983) work, since they only studied the axysimmetric case in which
the stars and the clump are aligned \footnote{The assumption made by
Klein et al, in which both stars are identical and located at the same
distance from the clump implies that the clump should not be accelerated
since it should actually be photo-evaporated at the same rate on both
sides faced to the stars. Also, this situation leads to a {\it minimum
shadow configuration} when the distances from the clump to the stars are
much greater than the clump radius: a situation that we will adopt in the
present paper.  Since one of our goals in the present work is to follow
the evolution of the shadow behind the clump as it is accelerated by the
rocket effect, and since that the axi-symmetric case has been already
treated in the literature, we will not address here the case of a aligned
system.}.  We first present an analytical description of the shape of the
shadow behind the clump considering both the distance from the sources
as well as their angular displacement\footnote{In a pioneering study,
\citet{dyson} found that a clump subject to an angular distribution
of ionizing radiation field should respond contracting its radius
and increasing its local density towards the maxima in the radiation
field. Our approach differs considerably from this paper since we also
follow the dynamical evolution of the clump and its associated shadow.}.
A set of three-dimensional numerical simulations is also presented,
in which we follow the evolution of the shadows as well as the clump
radius and neutral mass.

This paper is planned as follows. In \S 2, we present an analytical
solution for the shape of the shadows behind a clump illuminated
by two sources. In \S 3, we explore the parameters numerically and we
present the results of three-dimensional numerical simulations of this
problem. In \S 4, the discussions and conclusions are presented.

\section{Shadows behind illuminated clumps}
\label{sec:shadows}

\begin{figure}[!t]
  \includegraphics[width=\columnwidth]{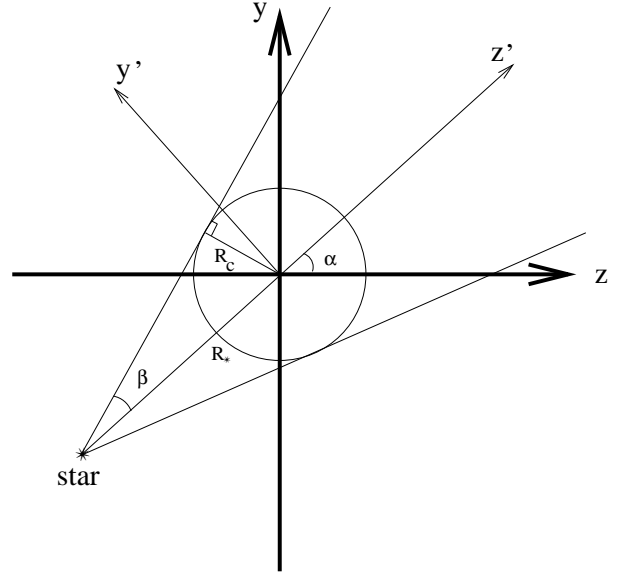}
  \caption{Schematic diagram showing i) the relation between the two
            coordinate systems, $O$ and $O'$, ii) the clump represented
            by a sphere of radius $R_c$ (projected onto the $yz-$plane),
            iii) the angles $\alpha$ and $\beta$ and iv) 
            the distance from the star $R_{\star}$. See the text
            for a discussion.}
  \label{F1}
\end{figure}

Figure \ref{F1} defines two frames of reference $O$ and $O^{\prime}$.
Both frames have their origins at the center of a spherical clump of
radius $R_c$, and the star is on the $yz-$ and $y^{\prime}z^{\prime}-$
planes. Frame $O$ has an arbitrary orientation, while frame $O^{\prime}$
is oriented such that its $z^{\prime}$-axis is along the line joining
the center of the clump with the star. The star is located at a distance
$R_{\star}$ from the clump and has coordinates: $x=0$, $y=-R_{\star}{\rm
sin}\alpha$ and $z=-R_{\star}{\rm cos}\alpha$ in frame $O$; and
$x^{\prime}=0$, $y^{\prime}=0$ and $z^{\prime}=-R_{\star}$ in frame
$O^{\prime}$ (see Figure \ref{F1}). The half opening angle $\beta$
of the shadow cone is given by:

\begin{figure}[!t]
  \includegraphics[width=\columnwidth]{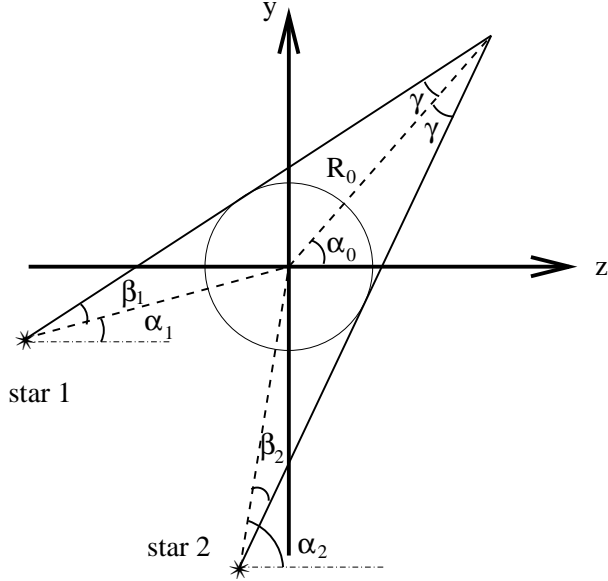}
  \caption{Schematic diagram showing the relative position, angles and
            distance between the two stars and the clump. See the text
            for a discussion.
            }
  \label{F2}
\end{figure}

\begin{equation}\label{eq1}
{\rm sin}\beta=\frac{R_c}{R_{\star}}.
\end{equation}

\noindent Thus the width $\omega$ of the shadow cone increases as,

\begin{equation}\label{eq2}
\omega={\rm tan}\beta(R_{\star}+z^{\prime})=
\frac{R_c}{(R_{\star}^2-R_c^2)^{1/2}}(R_{\star}+z^{\prime}).
\end{equation}

The shadow cone is tangent to the clump at,

\begin{equation}\label{eq3}
z_c^{\prime}=R_c{\rm sin}\beta,~~~ y_c^{\prime}=R_c{\rm cos}\beta,
\end{equation}

\noindent thus the shadow region behind the clump is given by,

\begin{equation}\label{eq4}
(x'^2+y'^2)^{1/2} \le \omega = {\rm tan}\beta(R_{\star}+z^{\prime}),
\end{equation}

\noindent and

\begin{equation}\label{eq5}
z^{\prime} \ge z_c^{\prime} = -R_c{\rm sin}\beta,
\end{equation}

\noindent where the equal sign in (\ref{eq4}) corresponds to the boundary
of the shadow.

The coordinate transformation equations between frames $O$ and
$O^{\prime}$ are,

\begin{eqnarray}\label{eq6}
 x^{\prime}=x, ~~~~~~~~~~~~~~~~~\nonumber \\
 y^{\prime}=y\,{\rm cos}\alpha - z\,{\rm sin}\alpha, \\
 z^{\prime}=y\,{\rm sin}\alpha + z\,{\rm cos}\alpha. \nonumber
\end{eqnarray}

Substituting (\ref{eq6}) in (\ref{eq4}) and (\ref{eq5}) we find that the
shadow region in frame $O$ is,

\begin{eqnarray}\label{eq7}
[x^2+(y\,{\rm cos}\alpha-z\,{\rm sin}\alpha)^2]^{1/2} \le ~~~~~~~~~~~~~~~~~~\nonumber \\
\frac{R_c}{(R_{\star}^2-R_c^2)^{1/2}}(R_{\star}+z\,{\rm cos}\alpha
+y\,{\rm sin}\alpha),
\end{eqnarray}

\noindent and

\begin{equation}\label{eq8}
y\,{\rm sin}\alpha+z\,{\rm cos}\alpha
\ge -\frac{R_c^2}{R_{\star}}.
\end{equation}

\begin{figure}[!t]
  \includegraphics[width=\columnwidth]{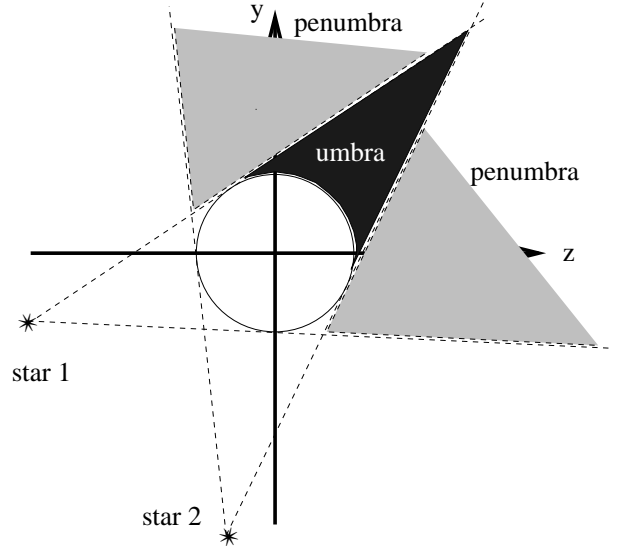}
  \caption{Schematic diagram showing the umbra and the penumbra regions
            in the shadow behind a clump illuminated by two stars.
            }
  \label{F3}
\end{figure}

Consider the case of a clump illuminated by two stars, one located at a
distance $R_1$ at an angle $\alpha_1$ and the other at a distance $R_2$
with an angle $\alpha_2$. The $yz-$plane contains both stars and the
center of the clump (see Figure \ref{F2}). The shadow produced by the
two stars is divided in two regions.  One, the {\it umbra} where the
light from both stars is blocked by the clump, and the other, the {\it
penumbra} where only the light from one of the stars is shaded by the
clump (see Figure \ref{F3}).

In the $yz-$plane ($x=0$) the boundary of the shadows are straight
lines given by (see eq. \ref{eq7}),

\begin{equation}\label{eq9}
y\,{\rm cos}\alpha_1 - z\,{\rm sin}\alpha_1 =
\pm {\rm tan}\beta_1 (R_1 +z\,{\rm cos}\alpha_1+
y\,{\rm sin}\alpha_1)
\end{equation}

\noindent for star 1, and,

\begin{equation}\label{eq10}
y\,{\rm cos}\alpha_2 - z\,{\rm sin}\alpha_2 =
\pm {\rm tan}\beta_2 (R_2 +z\,{\rm cos}\alpha_2+
y\,{\rm sin}\alpha_2)
\end{equation}

\noindent for star 2, where,

\begin{equation}\label{eq11}
{\rm tan}\beta_1=\frac{R_c}{(R_1^2-R_c^2)^{1/2}}
~~{\rm and}~~
{\rm tan}\beta_2=\frac{R_c}{(R_2^2-R_c^2)^{1/2}}{\rm .}
\end{equation}

Assuming that $\alpha_2 > \alpha_1$, then the maximum extent of the umbra
$R_0$ (see Figure \ref{F2}) is given by the intersection of lines
(\ref{eq9}) (with the plus sign) and line (\ref{eq10}) (with the
minus sign). The coordinates of the intersection point are,

\begin{equation}\label{eq12}
x_0=0, ~~ y_0=R_0{\rm sin}\alpha_0 ~~{\rm and}~~z_0=R_0{\rm cos}\alpha_0{\rm ,}
\end{equation}

\noindent where

\begin{equation}\label{eq13}
R_0=\frac{R_c}{{\rm sin}\gamma},
\end{equation}

\noindent and

\begin{equation}\label{eq14}
\alpha_0=\frac{(\alpha_1 + \alpha_2)-(\beta_2-\beta_1)}{2},
\end{equation}

\begin{equation}\label{eq15}
\gamma=\frac{(\alpha_2 - \alpha_1)-(\beta_1+\beta_2)}{2}.
\end{equation}

\noindent The geometrical interpretation of $\alpha_0$ and $\gamma$
is shown in Figure \ref{F2}.

It follows from (\ref{eq13}) that $R_0 \rightarrow
\infty$ and $\gamma \rightarrow 0$, for $(\alpha_2 -
\alpha_1)\rightarrow(\beta_1+\beta_2)$. Therefore for $(\alpha_2 -
\alpha_1) < (\beta_1+\beta_2)$ the umbra is not bounded and it extends
to infinity.

Let us define a function $J(x,y,z,\alpha,R_{\star})$ such
that $J=1$ if (\ref{eq7}) and (\ref{eq8}) are satisfied
and $J=0$ otherwise. Then, for the umbra $J(x,y,z,\alpha_1,R_1)=1$
and $J(x,y,z,\alpha_2,R_2)=1$ while for the penumbra either
$J(x,y,z,\alpha_1,R_1)=0$ and $J(x,y,z,\alpha_2,R_2)=1$
or $J(x,y,z,\alpha_1,R_1)=1$ and $J(x,y,z,\alpha_2,R_2)=0$. In
the region illuminated by both stars $J(x,y,z,\alpha_1,R_1)=0$
and $J(x,y,z,\alpha_2,R_2)=0$.

There is one case of particular interest, that in which both $R_1$ and
$R_2$ are much greater than $R_c$, and that we will explore in the next
section numerically. In the limit, $(R_c/R_1)$ and $(R_c/R_2)\rightarrow
0$, both $\beta_1$ and $\beta_2 \rightarrow 0$ and thus [see (\ref{eq14})
and (\ref{eq15})],

\begin{equation}\label{eq16}
\alpha_0 = \frac{(\alpha_1 + \alpha_2)}{2}~~
{\rm and}~~
\gamma=\frac{(\alpha_2 - \alpha_1)}{2}{\rm ,}
\end{equation}

\noindent with $R_0$ given by (\ref{eq13}).

To get an idea of the shape of the umbra in this case let us consider
a particular orientation of our frame of reference. This orientation is
such that

\begin{equation}\label{eq17}
\alpha_2=\pi -\alpha_1\,.
\end{equation}

\noindent Then, from (\ref{eq16}),

\begin{equation}\label{eq18}
\alpha_0=\frac{\pi}{2} ~~{\rm and}~~\gamma=\frac{\pi}{2}-\alpha_1,
\end{equation}

\noindent and from (\ref{eq13}),

\begin{equation}\label{eq19}
R_0=\frac{R_c}{{\rm cos}\alpha_1}.
\end{equation}

The boundary of the shadow projected by each star is [see (\ref{eq7})],

\begin{equation}\label{eq20}
x^2+(y\,{\rm cos}\alpha_1 - z\,{\rm sin}\alpha_1)^2=R_c^2
\end{equation}

\noindent for star 1, and,

\begin{equation}\label{eq21}
x^2+(y\,{\rm cos}\alpha_2 - z\,{\rm sin}\alpha_2)^2=R_c^2
\end{equation}

\begin{figure}[!t]
  \includegraphics[width=\columnwidth]{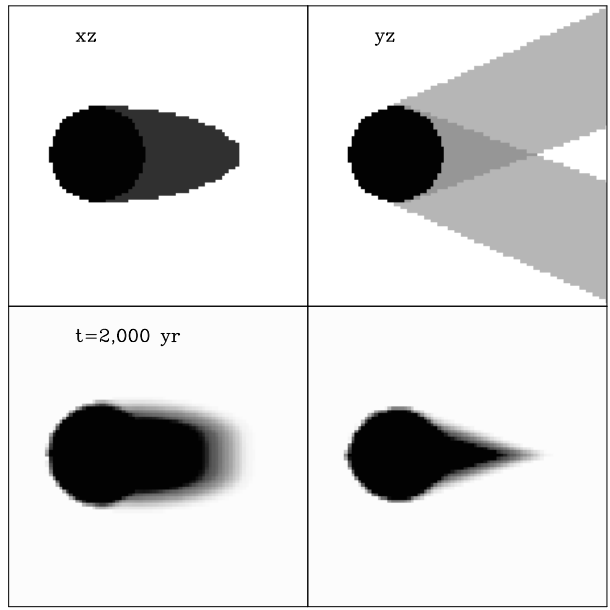}
  \caption{{\it Top:} A spherical clump and its projected umbra and
  penumbra on the $xz$- (left) and $yz$-planes
  (right), assuming that the two stars are on the $yz$-plane.
  {\it Bottom:} Maps of the ionization fraction at the beggining of the
  calculation ($t=2\,000$ years). As in the previous maps, a spherical
  clump creates a shadow on the $xz$- (left) and $yz$-planes (right).
  In all cases, the abscissa corresponds to the $z$-coordinate and
  is limited by $40 < z < 130$, and the ordinate corresponds to the $x$
  (left) or $y$ (right), both limited by $19 < x,y < 109$ (in units of 
  $7.8\times10^{15}$ cm).
            }
  \label{F4}
\end{figure}

\begin{figure}[!t]
  \includegraphics[width=\columnwidth]{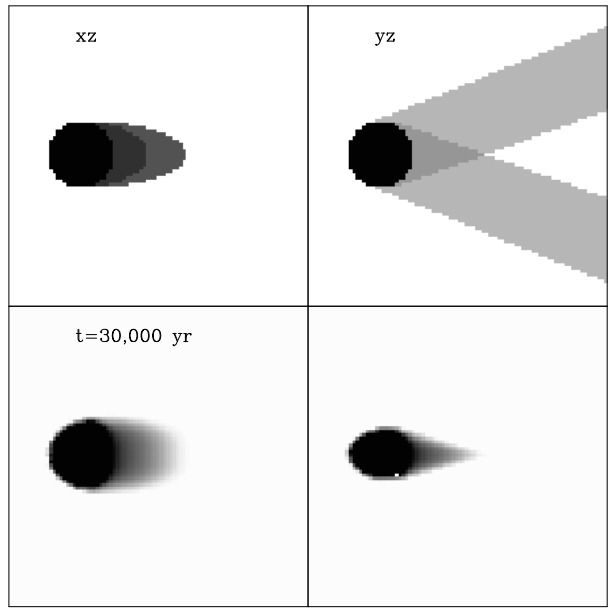}
  \caption{The same as in Figure \ref{F4}, but for a more evolved
  temporal stage ($t=30\,000$ years). In all cases, the abscissa
  corresponds to the $z$-coordinate and is limited by $80 < z < 170$
  (as the clump is accelerated towards the positive $z$ direction; see \S 3
  below), and the ordinate corresponds to the $x$ (left) or $y$ (right),
  both limited by $19 < x,y < 109$ (in units of $7.8\times10^{15}$ cm).
            }
  \label{F5}
\end{figure}

\noindent for star 2. Solving (\ref{eq20}) and (\ref{eq21}) simultaneously
and using (\ref{eq17}) it follows that the condition for the intersection
of the boundaries is $z=0$. That is, the shadows intersect at the
plane $z=0$, which is consistent with the value of $\alpha_0=\pi/2$
[see (\ref{eq18})]. In this plane, the shape of the umbra is then,

\begin{equation}\label{eq22}
x^2+y^2{\rm cos}^2\alpha_1 = R_c^2,
\end{equation}

\noindent which represents an ellipse centered at the center of the clump
with semi-axes $R_c$ in the $x$-direction and $R_0=R_c/{\rm cos}\alpha_1$
in the $y$-direction.

In order to compare the results from these analytical solutions with
those from the numerical simulations, let us anticipate here some results
before starting to discuss the simulations in detail in the next section.
Although we have explored different angular distributions of these two
stars with respect to the clump position, let us take here the case in
which $2\gamma=45^{\circ}$ (see Figure \ref{F2}), and a distance from
the clump to the two stars of $d=3.6\times 10^{18}$ cm. This situation
corresponds to model M45 discussed in the next section (see below).
In Figure \ref{F4}, we show both the analytical solution for the
shadow geometry (top) and the ionization fraction (which should trace the
shadow behind the clump) obtained through fully 3-D numerical simulations
(bottom) of model M45. Also, Figure \ref{F4} shows the shadows on the
$xz$- (left) and $yz$-plane (right). The stars and the clump are on the
$yz$-plane. We note that there is a good agreement between the predicted
geometry for the shadow (top-diagrams) compared with the ones obtained from
the simulations (bottom-diagrams). As the simulations progress in time,
the clump is photoevaporated and assumes a non-spherical shape, which
can be seen in the bottom-diagrams of Figure \ref{F5}. Such
a departure from a spherical shape produces small differences
between the predicted (see top-diagrams in Figure \ref{F5}) and
the simulated shadow geometry.

\section{The numerical simulations}

\subsection{The numerical method and the simulations}

\begin{figure*}[!t]
  \includegraphics[width=16cm]{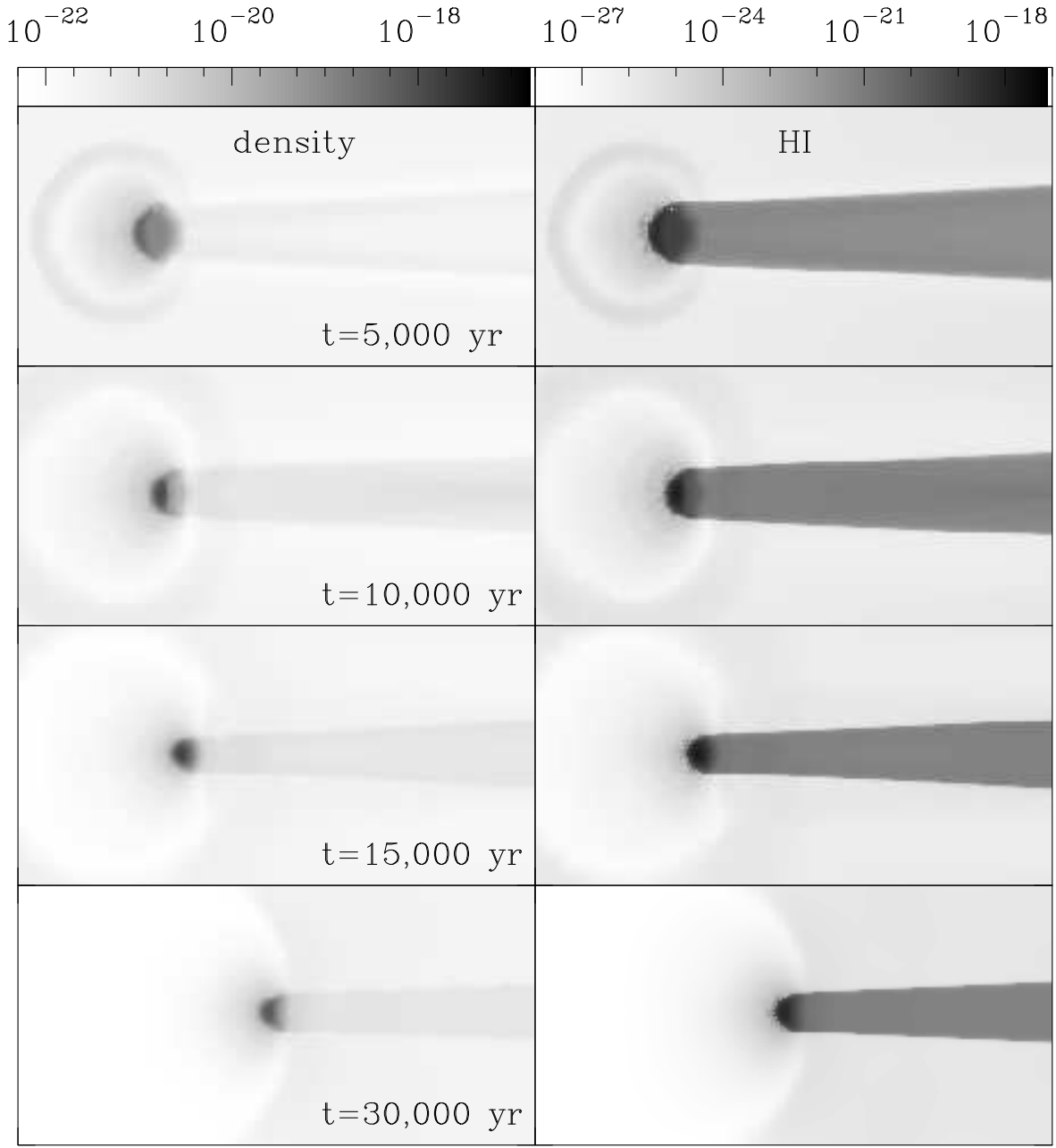}
  \caption{Distribution on the $yz-$plane of the total (left) and hydrogen neutral densities
            (right) for model M0 (see Table \ref{table1}) at $t=5,000$, 10,000,
            15,000 and 30,000 years (from top to bottom). The bars on the top
            give the scale in g cm$^{-3}$. The ionizing front comes
            from the left part of the computational domain (the ionizing
            sources are not depicted here).}
  \label{F6}
\end{figure*}

In order to investigate the temporal behavior of the shadows of
neutral (or partially neutral) material behind the clump exposed to
two ionizing sources, we have  carried out a set of three-dimensional
numerical simulations. The simulations were performed using the
Yguaz\'u-a code. The Yguaz\'u-a code, a binary adaptive grid code
\citep[see, e.g.,][]{raga00a, raga02}, has been extensively used
in the literature, and has been tested against analytical solutions
and laboratory experiments \citep[see, e.g.,][]{raga00a, sobral00,
raga01, vel01, raga04}. The Yguaz\'u-a code integrates the gas-dynamic
equations (employing the ``flux vector splitting'' scheme of Van Leer).
The code also solves rate equations for neutral/ionized hydrogen,
and the radiative transfer of the direct photons at the Lyman limit
\citep[see][]{gon04}.

We have computed models assuming a high density ($n_{c}=5\times10^4$), low
temperature clump ($T_{c}=10$ K) immersed in a low density ($n_{e}=100$
cm $^{-3}$), high temperature environment ($T_{e}=1000$ K). All the models
assume a clump of radius $r_{c}=10^{17}$ cm located at 1 parsec from the
ionizing sources. The ionizing sources were assumed to have a black-body
spectrum with an effective temperature of $T_{\star}={5\times 10^4}$ K
and an emission rate of ionizing photons of $S_{\star}=1.5\times10^{49}$
s$^{-1}$. In all models, the computational domain is limited to $x$,
$y=10^{18}$ cm, $z=2\times 10^{18}$ cm and a 5-level binary adaptative
grid with a maximum resolution of $7.81\times10^{15}$ cm along each
axis has been used.  The centre of the clump is initially located at
$x=y=z=5\times10^{17}$ cm.

\begin{table}[!t]\centering
  \setlength{\tabnotewidth}{\columnwidth}
  \tablecols{3}
  % Stretch the space between table columns
  \setlength{\tabcolsep}{1.8\tabcolsep}
  \caption{The simulated models} \label{table1}
  \begin{tabular}{lrrr}
    \toprule
Model & $\theta$ ($^{\circ}$) & $S_1$ ($s^{-1}$) & $S_2$ ($s^{-1}$) \\
    \midrule
M0   & 0  &  $1.50\times10^{49}$     & $1.50\times10^{49}$ \\
M45  & 45 &  $1.50\times10^{49}$     & $1.50\times10^{49}$ \\
M45b & 45 &  $1.50\times10^{49}$     & $1.50\times10^{48}$ \\
M90  & 90 &  $1.50\times10^{49}$     & $1.50\times10^{49}$ \\
    \bottomrule
  \end{tabular}
\end{table}

In order to explore the geometrical effect on the shadows due to the
presence of two ionizing sources, we have computed models with distinct
relative angles between the clump and the two sources.  In Table
\ref{table1} we give the angles $\theta$, which are equivalent to
$2\gamma$ as previously defined in Figure \ref{F2} (see also equations
\ref{eq15} and \ref{eq16}).  In particular $\theta=0$ for model M0,
$\theta=45^{\circ}$ for model M45 and $\theta=90^{\circ}$ for model
M90 (see Table \ref{table1}). We also note that we have conducted
simulations with different ratios between the ionizing fluxes from
the two sources, $F=S_{\star,1}/S_{\star,2}$, namely, $F=1$, and 10
(with $S_{\star,1}=1.5\times10^{49}$ s$^{-1}$) . In the next section,
we will discuss in detail the results from these numerical experiments.

\subsection{Results}

\begin{figure*}[!t]
  \includegraphics[width=16cm]{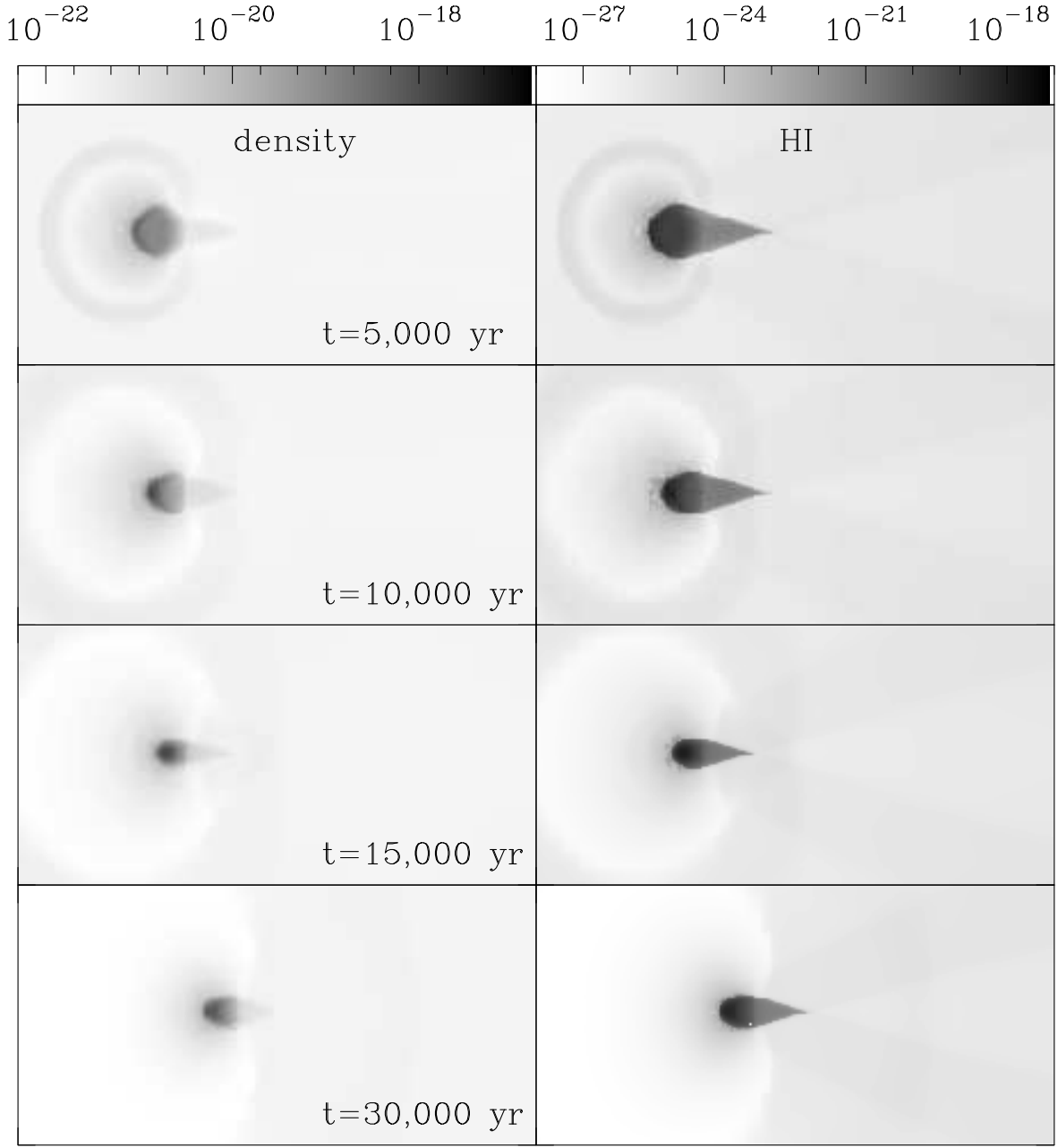}
  \caption{The same as in Figure \ref{F6}, but for model M45
    (see Table \ref{table1}).}
  \label{F7}
\end{figure*}

\begin{figure*}[!t]
  \includegraphics[width=16cm]{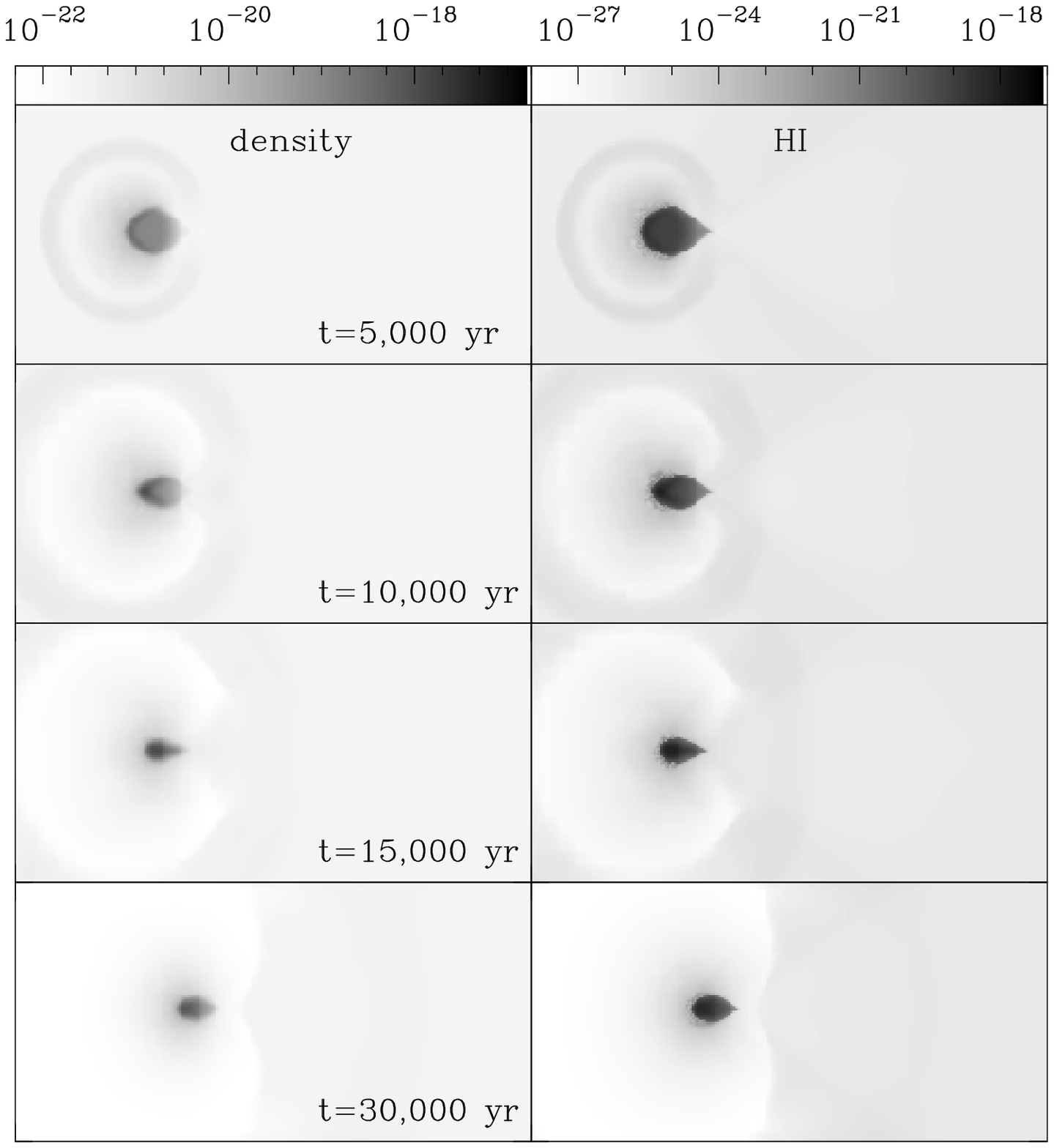}
  \caption{The same as in Figure \ref{F6}, but for model M90
    (see Table \ref{table1}).}
  \label{F8}
\end{figure*}

\begin{figure*}[!t]
  \includegraphics[width=16cm]{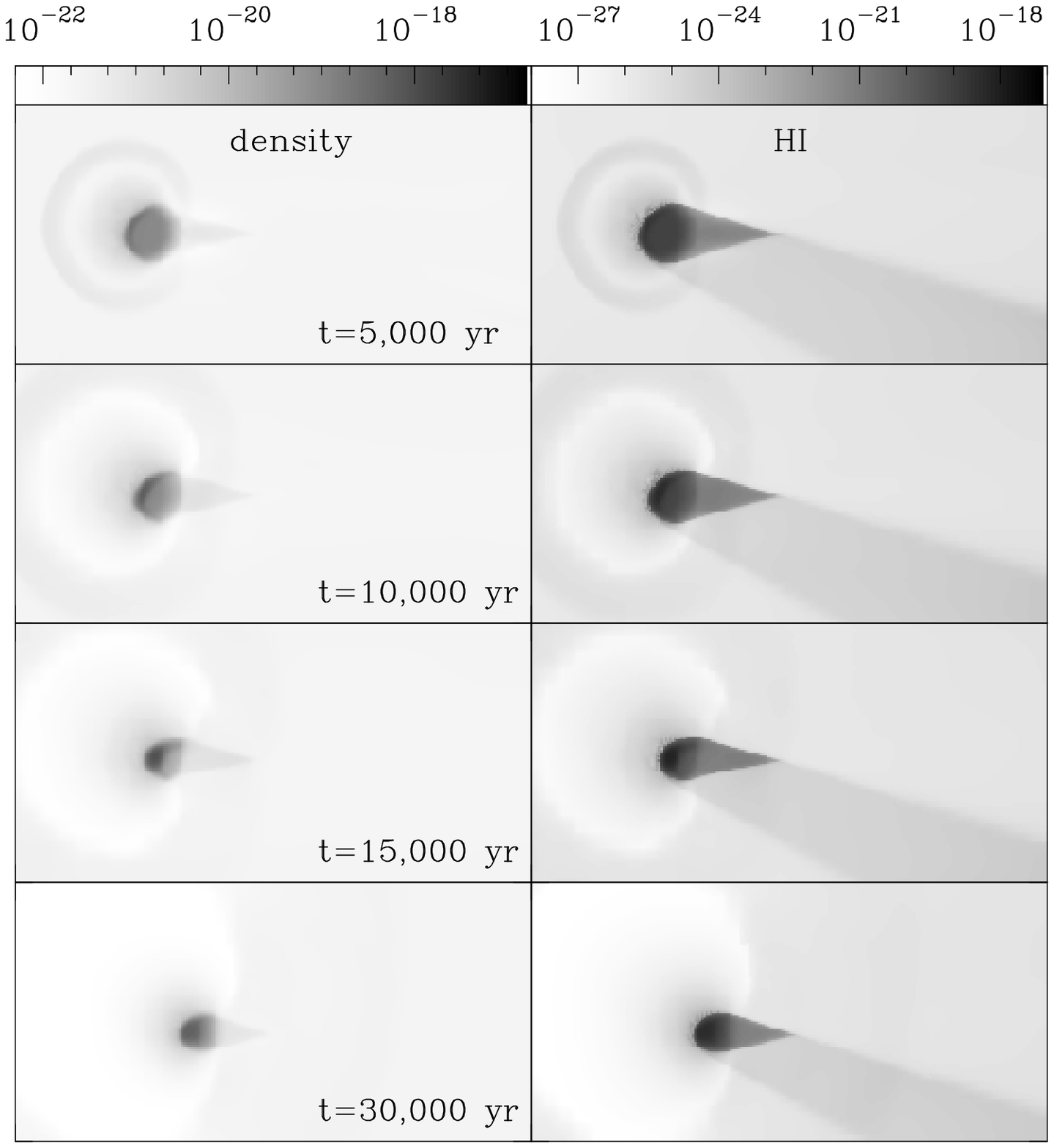}
  \caption{The same as in Figure \ref{F6}, but for model M45b
   (see Table \ref{table1}).}
  \label{F9}
\end{figure*}

Figures \ref{F6}, \ref{F7} and \ref{F8} show the temporal evolution
of the total density (left) and the neutral hydrogen density (HI; right)
for models M0, M45 and M90, respectively  (see Table \ref{table1}) at
$t=5,000$, 10,000, 15,000 and 30,000 years. In these density distribution
maps, we can identify {\it i)} a tail that fills the shadow region behind
the clump (the umbra in Figure \ref{F3}), {\it ii)} the emergence of a
photoevaporated flow (that propagates towards the ionizing sources), and
{\it iii)} the acceleration of the clump due to the {\it rocket-effect}.

As discussed in \S 2, the umbra is expected to have an extension
$R_0$, which is in turn controlled by two parameters, namely, the
clump radius $R_c$ and the relative angle $\gamma$ with respect to the
sources (see equation \ref{eq13} and Figure \ref{F2}). In particular,
the simulations show that the tail which appears in the shadow region
behind the clumps has a cylindrical shape for model M0 ($R_0 \rightarrow
\infty$; Fig. \ref{F6}), while in models M45 and M90 the umbra depicts
a conical shape (see Figures \ref{F7} and \ref{F8}). The conical
umbra, however, is somewhat smaller in model M90 when compared with
the one of model M45. We also note that the the cone height $R_0$ is
actually a function of time ($R_0=R(t)$), since the photoevaporation of
the clump material reduces the clump radius (see the discussion below).
In all cases, both the umbra and the penumbra are completely symmetric
with respect to the $z$-axis.

The photoevaporated flow can be seen in both the total density and HI
density maps in Figures \ref{F6}, \ref{F7} and \ref{F8}. It expands
almost radially from the clump position with a velocity $\lesssim 20$
km s$^{-1}$, producing density perturbations in the environment (a shock
wave). This photoevaporated flow accelerates outwards, towards the ionizing
photon sources.  As a consequence, the clump itself is accelerated in the
opposite direction. This effect is clearly seen in Figures \ref{F6},
\ref{F7} and \ref{F8}, where the clump is continuously pushed towards
larges distances from the ionizing sources. However, we also note that
the mass flux in the photoevaporated flow is smaller in model M90 than
in model M45, and, as a consequence, the acceleration of the clump is
also lower (see the discussion below).

In Figure \ref{F9} we show the evolution of the total density
(left) and the neutral hydrogen density (HI; right) for model M45b
(see Table \ref{table1}) at $t=5,000$, 10,000, 15,000 and 30,000 years.
We note that in this particular model, in which the ionizing sources have
different ionizing photon rates ($S_{\star,1}/S_{\star,2}=10$), both the
photoevaporated flow and the shadows behind the clump, particularly the
penumbra, are not symmetric with respect to the $x$-axis. In particular,
the photoevaporated flow is stronger in the direction of the most powerful
ionizing source, $S_1$ (located towards the top-left direction of the
diagram; see Figure \ref{F9}). This causes a more pronounced ablation of
the clump material that is facing the more powerful ionizing source. Also,
only the penumbra of the shadow produced by the ionizing field from $S_1$
is seen.

\begin{figure}[!t]
  \includegraphics[width=\columnwidth]{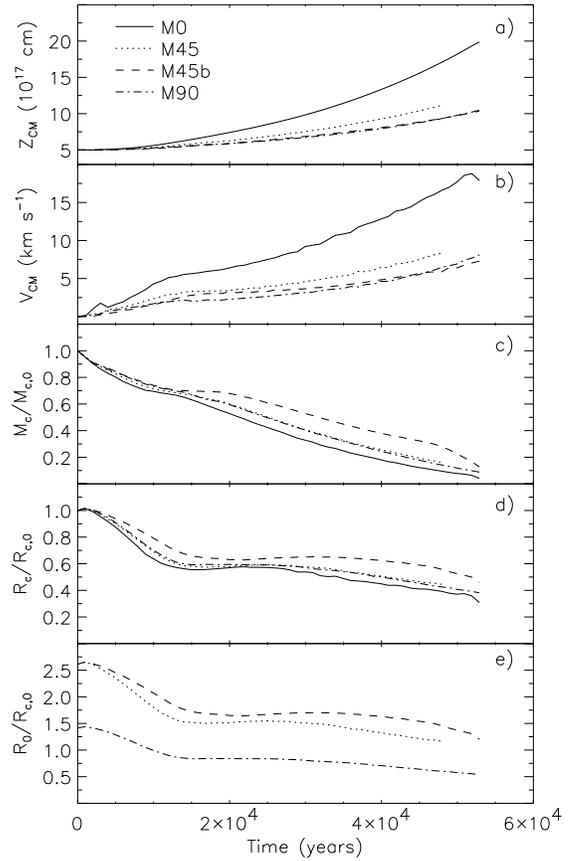}
  \caption{{\it a)} Clump centre of mass (CM) position
    (measured by $Z_{CM}$, in units of $10^{17}$ cm) as a function of
    time for models M0 (full line), M45 (dotted line), M45b (dashed line)
    and M90 (dot-dashed line) (see Table \ref{table1}). We note that the
    variations in $X_{CM}$ and $Y_{CM}$ are almost null for all models
    with the exception of M45b, see the discussion in the text.
    {\it b)} Velocity of the clump centre of mass ($V_{CM}$; in units
    of km s$^{-1}$) as a function of time (the curves are labeled in
    the top panel).
    {\it c)} Clump mass (normalized by the initial clump mass,
    $M_{c,0}=4.55\times10^{32}$ g) as a function of time (the curves
    are labeled in the top panel).
    {\it d)} Clump radius (normalized by the initial clump radius,
    $R_{c,0}=10^{17}$ cm) as a function of time (the curves are labeled
    in the top panel).
    {\it e)} Shadow parameter $R_0$ (see equation \ref{eq13}), normalized
    by the initial clump radius, as a function of time (we note that
    $R_0 \rightarrow \infty$ for model M0; not shown here).}
  \label{F10}
\end{figure}

In Figures \ref{F10}a and \ref{F10}b we show, respectively, the
position $Z_{CM}$ of the center of mass of the clump and its velocity
as a function of time (in years), for models M0 (full line), M45 (dotted
line), M45b (dashed line) and M90 (dot-dashed line)\footnote{We note that
the coordinates $X_{CM}$ and $Y_{CM}$ are almost constant, and equal
to the initial value $X_{CM}(t=0)=Y_{CM}(t=0)=5\times 10^{17}$ cm, for
all models with the exception of model M45b, where $Y_{CM}$ changes with
time. Hence, the match in the curves from model M45b (dashed line) and M90
(dot-dashed line) in this diagram does not mean that the clump has the
same velocities.}. As mentioned before, the clump in model M0 is rapidly
accelerated (when compared with the clump in the other models) and attains
large distances from the ionizing sources at shorter times. This is due
to the fact that the mass loss rate associated with the photoevaporated
flow, $\dot{M_c}$, is higher in this case. This is illustrated in Figure
\ref{F10}c, where the clump mass (normalized by the initial clump mass,
$M_c(t=0)=M_{c,0}=\rho_0(4\pi/3)R_{c,0}^3= 4.55\times10^{32}$ g) is
shown for all models, as a function of time. Note that the clump mass
loss in the M0 model is higher (when compared with the other models).

\begin{figure}[!t]
  \includegraphics[width=\columnwidth]{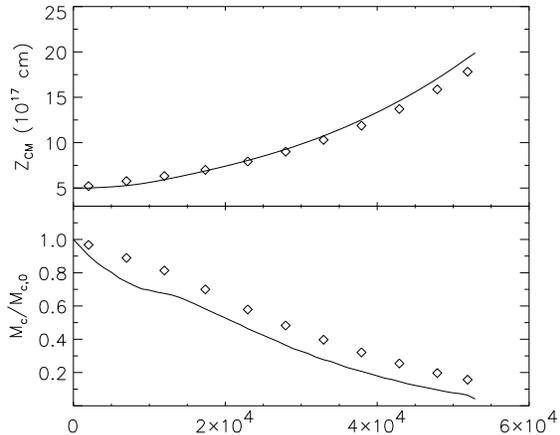}
  \caption{{\it Top:} Clump centre of mass (CM) position
    (measured by $Z_{CM}$, in units of $10^{17}$ cm) as a function
    of time for simulated model M0 (full line; the same as in Figure
    \ref{F10}).  Also depicted in this figure is the result from the
    analytical approximation (diamonds) from \citet{mellema98}, using
    as input parameters the sound speed $c_0$ and $c_i$ taken from the
    numerical simulation (see the text).
    {\it Bottom:} Clump mass (normalized by the initial clump mass,
    $M_{c,0}=4.55\times10^{32}$ g) as a function of time taken from the
    M0 model (full line; the same as in Figure \ref{F10}) . As in the
    previous diagram, the diamonds represents the analytical solution.
    }
  \label{F11}
\end{figure}

In Figure \ref{F10}d we also show the spherical clump radius $R_c$
(normalized by the initial clump radius, $R_c(t=0) = R_{c,0} = 10^{17}$
cm), as a function of time, for all of the models listed in Table
\ref{table1}. All of the curves show an almost monotonic behaviour. Note
that the clump radius in model M0 is always smaller than the values
obtained from the other models (which is consistent with the higher
$\dot{M_c}$ of model M0). The temporal variation of the clump radius is
responsible for the temporal variation of the height of the conical shadow
$R_0$, as can be seen in figures \ref{F7}, \ref{F8} and \ref{F9}. In
Fig. \ref{F10}e, the height of the conical shadow $R_0$ (normalized
by the initial clump radius, $R_{c,0}$; $R_0 \rightarrow \infty$ for
model M0, not shown in Figure \ref{F10}e) is calculated\footnote{These
curves show the qualitatively behavior of $R_0$ with time, since we
are calculating it at each time step, without taking into consideration
the adjustment of the shadow to the steady-state.}  using the clump
radius given in Fig. \ref{F10}d and equation (\ref{eq13}). Note that,
after a strong variation at the beginning ($t \lesssim 10^4$ years),
the conical height tends to values $\approx R_0(t=0)/2$ in all models.

We have also compared the results from our simulations with results
obtained from the analytical approximations given by \citet{mellema98}.
In that paper, the authors studied both the collapse (or implosion) and
the cometary phases of a clump being exposed to an incident plane-parallel
ionizing front (the case of model M0 here), and the mass and the position
of the clump as a function of time have been derived for both phases.
Using values obtained from our simulations for model M0 (following the
notation in Mellema et al.: sound speed at the base of the photoevaporated
flow, $c_i \simeq 8.64$ km s$^{-1}$; sound speed of the shocked gas
inside the clump, $c_0 \simeq 2.12$ km s$^{-1}$), we found that the
collapse phase takes $t_0 = 13\,234$ years. Interestingly, almost all the
curves in Figure \ref{F10} seems to change their behavior at this time,
and this seems to be due to the end of this phase and the beginning of
the cometary phase. At time $t_0$, the clump is predicted to have a mass
$M_c \simeq 0.797\times M_{c,0}$. Figure \ref{F10}c shows that model M0
has a mass at this time that is $\sim$ 10\% smaller than this value,
in good agreement with the analytic theory.  Finally, the behavior of
the solid curves in Figure \ref{F10}a,c has also been compared with
the solutions presented by \citet{mellema98} and a very good agreement
is found between the values from the simulations and those predicted by
the analytic theory during all of the evolution of the clump in model
M0. 

In particular, in Figure \ref{F11} we plot again the result from
model M0 (full lines) and its comparison with the result from the
analytical approximation from \citet{mellema98} (diamonds). Although
the clump CM position is well recovered in the simulation (top panel),
it seems that the mass of the clump is under-estimated in the numerical
simulation when compared with the analytical solution. However, even in
this case, both curves (full line and diamonds, on the bottom panel of
Figure \ref{F11}) depicts the same behavior. We have also simulated
this same model M0, but with a grid-resolution of 512x128x128 (these
numbers corresponding to a uniform grid at the highest resolution of our
adaptative grid), or a grid spacing of $3.9\times10^{15}$ cm (i.~e., an
improvement of a factor 2 with respect to the simulations that we have
presented up to now).  As in the previous, low resolution case, the CM
position is well reproduced by the numerical simulation (not shown here),
and the same under-estimation of the clump mass is also obtained.

\section{Conclusions}

In this paper, we have explored analytically what is the shape of the
shadow behind a clump exposed to two ionizing photon sources.  For the
case in which the clump radius is much smaller than the distances to
the sources, an analytical solution for the shapes of the umbra and the
penumbra are found. We have also carried out 3D numerical simulations
of this scenario, and we present four models with different relative
positions between the two stars and the clump as well as with different
combinations of ionizing photon rates for the sources.

These models show that the long neutral tails produced in the interaction
of a neutral clump with a single photoionizing source (model M0, see
Table 1 and Figure 4) disappear for the case of two angularly separated
photoionizing sources producing comparable ionizing photon fluxes at the
position of the clump (models M45 and M90, see Figures 5 and 6). This
shortening of the neutral tails is again obtained in a model in which
the ionizing photon fluxes from the two sources differ by a factor of 10
(model M45b, see Table 1 and Figure 7). In our numerical simulations, we
also obtain the time-evolution of the cometary clump as it is accelerated
away from the photoionizing sources by the rocket effect.

The present models are meant as an illustration of the effects of the
interaction of a neutral clump with two photoionizing sources, which might
be applicable to neutral structures in H~II regions with several massive
stars \citep[see, e.g.,][]{odell,smith}. However, in our simulations
we have not included the diffuse ionizing radiation (which might have
an important effect on the shadow region, in particular in the case of
ionization bounded nebulae) and the effect of the dissociative, direct
and diffuse radiation \citep[which can affect both the head and the tail
regions, see][]{rich}. In order to model the structure of specific clumps,
these effects should also be included.

\acknowledgements

We would like to thank the anonymous referee for the suggestions,
which improved substantially the presentation of the paper.
A.H.C. acknowledge Brazilian agency CAPES for a post-doctoral
fellowship (process BEX 0285/05-6). A.H.C. and M.J.V. would like to
thank P. Vel\'azquez and the staff of the ICN-UNAM (M\'exico), for
their warm hospitality and also for partial financial support during
our visiting. This work is funded in part by the Brazilian agencies
PROPP-UESC (220.1300.327) and CNPq (62.0053/01-1-PADCT III/Mil\^enio,
470185/2003-1 and 306843/2004-8). This work was also supported by
CONACYT grants 41320-F, 43103-F and 46828 and the DGAPA (UNAM) project
IN~113605.

{}

\end{document}